\documentclass[reprint,superscriptaddress,amsmath,amssymb,aps,pra,floatfix]{revtex4-2}

\usepackage[english]{babel}
\usepackage{microtype}
\usepackage[T1]{fontenc}
\usepackage{graphicx}
\usepackage{dcolumn}
\usepackage{bm}
\usepackage{mathdots}
\usepackage{color}
\usepackage{xcolor}
\usepackage{physics}
\usepackage[normalem]{ulem}
\usepackage{stmaryrd}
\usepackage{hyperref}
\usepackage{breqn}
\usepackage{cancel}
\usepackage{setspace}
\usepackage[titletoc]{appendix}
\usepackage{caption}
\usepackage{ragged2e}

\renewcommand{\vec}[1]{\mathbf{#1}}

\makeatletter
\let\cat@comma@active\@empty
\makeatother

\begin{document}

\preprint{APS/123-QED}

\title{{Nonlocal wavefront shaping through complex media}}

\author{Yanis Trouyet}
\email{yanis.trouyet@insp.upmc.fr}
\affiliation{Sorbonne Université, CNRS, Institut des NanoSciences de Paris, INSP, F-75005 Paris, France}

\author{Neelan Gounden}
\affiliation{University of the Witwatersrand, 9 Enoch Sontonga Ave, Johannesburg, South Africa}

\author{Pedro Ornelas}
\affiliation{University of the Witwatersrand, 9 Enoch Sontonga Ave, Johannesburg, South Africa}
 
\author{Patrick Cameron}
\affiliation{Dipartimento di Fisica, Università degli Studi di Napoli Federico II, Complesso Universitario di Monte Sant’Angelo, Via Cintia, 80126 Napoli, Italy}
\affiliation{Scuola Superiore Meridionale, Via Mezzocannone, 4, 80138 Napoli, Italy}

\author{Andrew Forbes}
\affiliation{University of the Witwatersrand, 9 Enoch Sontonga Ave, Johannesburg, South Africa}

\author{Hugo Defienne}
\email{hugo.defienne@insp.upmc.fr}
\affiliation{Sorbonne Université, CNRS, Institut des NanoSciences de Paris, INSP, F-75005 Paris, France}

\begin{abstract}
Wavefront shaping is a key technique for mitigating scattering in complex media, enabling advanced imaging and optical communication.
Yet existing approaches are inherently local, requiring active correction elements - such as spatial light modulators or deformable mirrors - to lie directly in the optical path of the scattered light, which limits their integration into compact imaging systems. 
Here, we experimentally demonstrate nonlocal wavefront shaping using spatially entangled photon pairs. 
By applying a phase correction to a photon that never interacts with the scattering medium, we compensate for the distortions experienced by its entangled partner and restore their initial spatial correlations. 
Our approach physically decouples the wavefront correction from the scattering medium, paving the way for imaging through complex media in compact and otherwise inaccessible systems.
\end{abstract}
\maketitle

When classical coherent light propagates through complex scattering media, the scrambling of amplitude and phase produces a speckle pattern. This severely limits optical applications, particularly deep-tissue imaging~\cite{bertolotti_imaging_2022}. To overcome this limitation, wavefront shaping (WS) techniques~\cite{cao_shaping_2022} use spatial light modulators (SLMs) to precisely tailor the optical field, compensating for scattering-induced distortions and restoring a diffraction-limited focus at the target plane~\cite{vellekoop_focusing_2007,popoff_measuring_2010}.

In recent years, these WS techniques have been adapted to the quantum regime~\cite{lib_quantum_2022}, particularly for the control of two-photon states. As in the classical case, the SLM is typically positioned within the optical path of the photon pairs to collectively shape their wavefront prior to entering the scattering medium. By employing auxiliary classical sources or through direct optimization of coincidence counts, iterative or matrix-based approaches enable the programming of the SLM to control quantum interference~\cite{defienne_two-photon_2016,wolterink_programmable_2016,leedumrongwatthanakun_programmable_2020, makowski_large_2024}, restore spatial correlations~\cite{defienne_adaptive_2018,lib_real-time_2020,devaux_restoring_2023,bajar_rapid_2025,courme_non-classical_2026,aarav_wavefront_2026,shekel_shaping_2024,shekel_shaping_2021}, recover entanglement at the output~\cite{courme_manipulation_2023} or even transmit sharp images~\cite{cameron_adaptive_2024,verniere_entanglement-enabled_2026}.

By exploiting the nonlocal nature of entanglement, one can manipulate the properties of a single photon to control the joint state of the entire pair~\cite{zhang_entanglement-based_2024}. This enables remote-controlled capabilities for quantum communication~\cite{ekert_quantum_1991}, computing~\cite{wang_remote-controlled_2020}, sensing~\cite{brown_interferometric_2023,stas_entanglement-assisted_2026}, and imaging~\cite{defienne_polarization_2021,ortolano_quantum_2023}. 
Recently, this dual-path approach has been extended to nonlocal aberration correction. 
For example, Black et al.~\cite{PhysRevLett.123.143603} restored spatial correlations degraded by a defocus present in one photon's path by displaying the quadratic phase pattern on an SLM in the path of its partner.
Similarly, Herrera et al.~\cite{valencia_unscrambling_2020} unscrambled a photon propagating through a multimode fiber by measuring its transmission matrix and applying targeted SLM phase corrections, successfully restoring spatial correlations one mode pair at a time to ultimately certify entanglement.

In this work, we demonstrate an iterative WS approach to restore the spatial correlations of entangled photon pairs when one photon propagates through a scattering medium while wavefront correction is applied to its partner. 
We present experimental results for both low-order aberrations, using a polydimethylsiloxane (PDMS) layer, and the single-scattering regime, using a Parafilm layer. 
We further validate the approach through numerical simulations in the multiple-scattering regime, using an experimentally measured transmission matrix of a thick stack of Parafilm layers. 
Finally, we analyze the achieved enhancement factor across these different scattering regimes and compare it with theoretical predictions.

\begin{figure}[h!]
    \centering
    \includegraphics[width=1\linewidth]{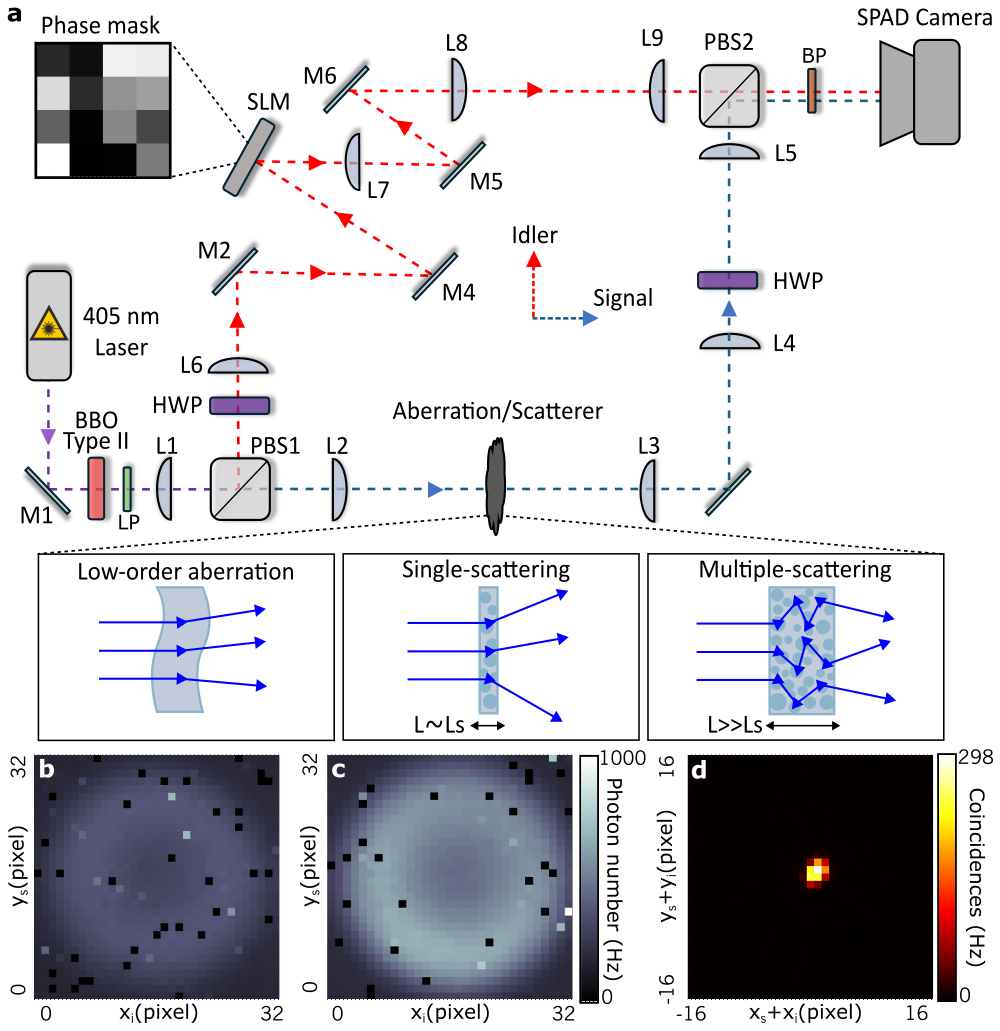}
   \caption{\textbf{Experimental setup.} \textbf{(a)} Spatially entangled photon pairs (810~nm) are generated via type-II spontaneous parametric down-conversion (SPDC) using a collimated 405~nm continuous-wave laser and a 1-mm $\beta$-barium borate (BBO) crystal. A 650~nm long-pass (LP) filter rejects the pump, and a polarizing beam splitter (PBS1) separates signal and idler photons. In the idler arm, a $4f$ system ($L_1=40$~mm, $L_2=200$~mm) images the crystal onto a spatial light modulator (SLM), which is Fourier-imaged onto a SPAD camera via a $4f$ setup ($L_6=100$~mm, $L_7=100$~mm) combined with lens $L_8=100$~mm. 
   In the signal arm, $L_1$ and $L_2$ image the crystal onto an aberrating or scattering medium (PDMS or Parafilm). 
   This plane is Fourier-imaged to the camera by a $4f$ system ($L_3= L_4=50$~mm) and lens $L_5=100$~mm. An $810 \pm 5$~nm bandpass (BP) filter removes non-degenerate pairs. 
   half-wave plates (HWPs) set at $45^\circ$ in each arm and PBS2 enable side-by-side imaging on the same sensor (M: mirror; L: lens). 
   \textbf{(b,c)} Aberration-free intensity images of the idler and signal regions (hot pixels set to zero, shown in black). 
   \textbf{(d)} Aberration-free $G^{(2)}$ sum-coordinate projection; the sharp peak indicates strong spatial anti-correlations.}   
       \label{Figure1}
\end{figure}

\textit{Experiment ---}The experimental setup is shown in Figure~\ref{Figure1}a. Spatially entangled photon pairs with orthogonal polarizations are generated via spontaneous parametric down-conversion (SPDC) and separated into two distinct optical arms, denoted idler (i) and signal (s). In each arm, lenses are arranged in a Fourier configuration to perform measurements in the momentum basis. At the output, the pairs are detected by two separate regions of a single-photon avalanche diode (SPAD) camera (denoted idler and signal regions). This enables the measurement of the spatial intensity correlation matrix $G^{(2)}(\vec{r}_i,\vec{r}_s)$, where $\vec{r}_i=(x_i,y_i)$ and $\vec{r}_s=(x_s,y_s)$ denote the pixel coordinates on each respective region of the sensor~\cite{ndagano_imaging_2020}.

Without optical aberrations, the idler and signal intensity images (Figs.~\ref{Figure1}b,c) reveal a characteristic SPDC ring structure. 
Furthermore, the $G^{(2)}$ sum-coordinate projection (Fig.~\ref{Figure1}d) exhibits a sharp central peak, showing strong spatial anti-correlations between the photon pairs measured at the output~\cite{moreau_realization_2012,edgar_imaging_2012,defienne_general_2018}. 
The metric $C_0$ is defined as this projection's central value ($\vec{r}_s+\vec{r}_i = 0$). 
Details on the $G^{(2)}$ measurement and projection are provided in Section~I of the Supplemental Material.

\begin{figure}[h!]
    \centering
    \includegraphics[width=1\linewidth]{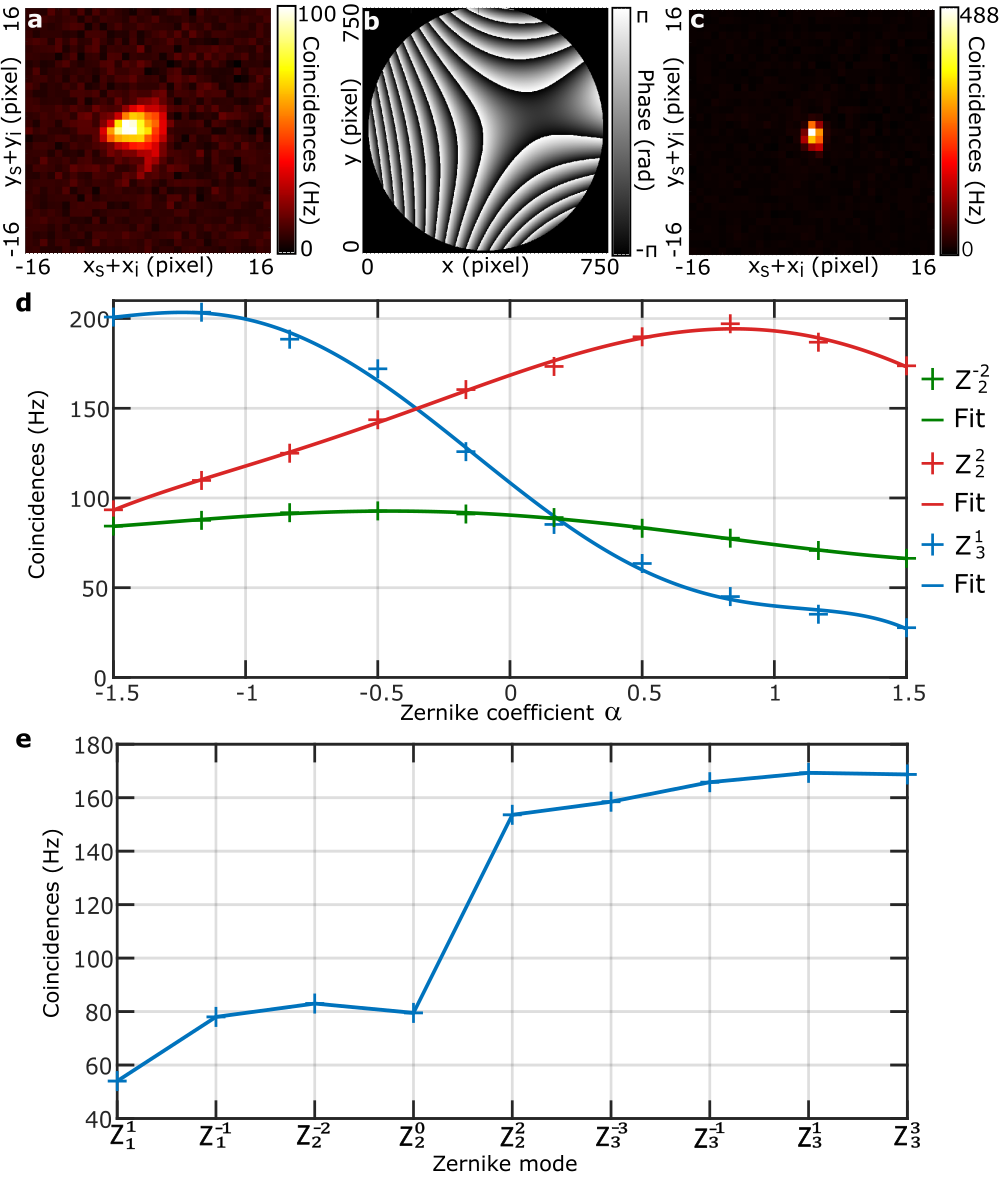}
    \caption{\textbf{Experimental results for low-order aberrations.} \textbf{(a)} Sum-coordinate projection of $G^{(2)}$ acquired in the presence of the PDMS layer with a flat phase applied to the SLM. \textbf{(b)} Final optimized phase mask. \textbf{(c)} Sum-coordinate projection of $G^{(2)}$ acquired with the optimized phase mask displayed on the SLM, demonstrating the restoration of strong spatial correlations between the photon pairs. \textbf{(d)} Correlation value $C_0$ as a function of the Zernike coefficient $\alpha_n^m$ for three Zernike modes: $Z_2^{-2}$ (green), $Z_2^{2}$ (red), and $Z_3^{1}$ (blue), along with their corresponding polynomial fits. \textbf{(e)} Optimization curve showing the evolution of $C_0$ as a function of the number of corrected Zernike modes on the SLM. Each $G^{(2)}$ measurement required the acquisition of $3.9.10^7$ frames at $1~\mu$s exposure.}
       \label{Figure2}
\end{figure}

\textit{Aberration regime ---} Inserting a PDMS layer into the signal arm near a reciprocal plane (conjugate to the crystal) introduces low-order aberrations, distorting the $G^{(2)}$ sum-coordinate peak and reducing its maximum intensity (Fig.~\ref{Figure2}a).  
To compensate, a Zernike-based adaptive optics algorithm~\cite{AOmicro} is implemented using an SLM in the idler arm, also positioned close to a reciprocal plane to the camera. 
This optimization yields a final phase mask -- a linear combination of Zernike polynomials with optimal coefficients (Fig.~\ref{Figure2}b) -- that successfully restores a sharp correlation peak (Fig.~\ref{Figure2}c). 
This optimal mask is determined through a sequential process: at iteration step $k$, the SLM displays phase masks $\theta_k(x,y) = \theta_{k-1}(x,y) + \alpha_n^m Z_n^m(x,y)$ across ten Zernike coefficient values $\alpha_n^m$, where $Z_n^m(x,y)$ is the current Zernike mode and $\theta_{k-1}(x,y)$ is the phase accumulated from previous steps. 
For each $\alpha_n^m$, we measure the correlation value $C_0$ and use a polynomial fit to extract the coefficient $\alpha_n^m[\text{max}]$ that maximizes this metric. This routine is repeated for all modes up to $n=3, m=3$ (Figs.~\ref{Figure2}d and e). 
Correction efficiency is quantified by the Strehl ratio $s$~\cite{mahajan_strehl_1983}, defined as the ratio of the final $C_0$ to its diffraction-limited value measured in Fig.~\ref{Figure1}d. 
We measure $s=1.75>1$, demonstrating that our approach corrects both the PDMS-induced aberrations and those inherent to the initial imaging system, which was not perfectly diffraction-limited.

\begin{figure}[h!]
    \centering
    \includegraphics[width=1\linewidth]{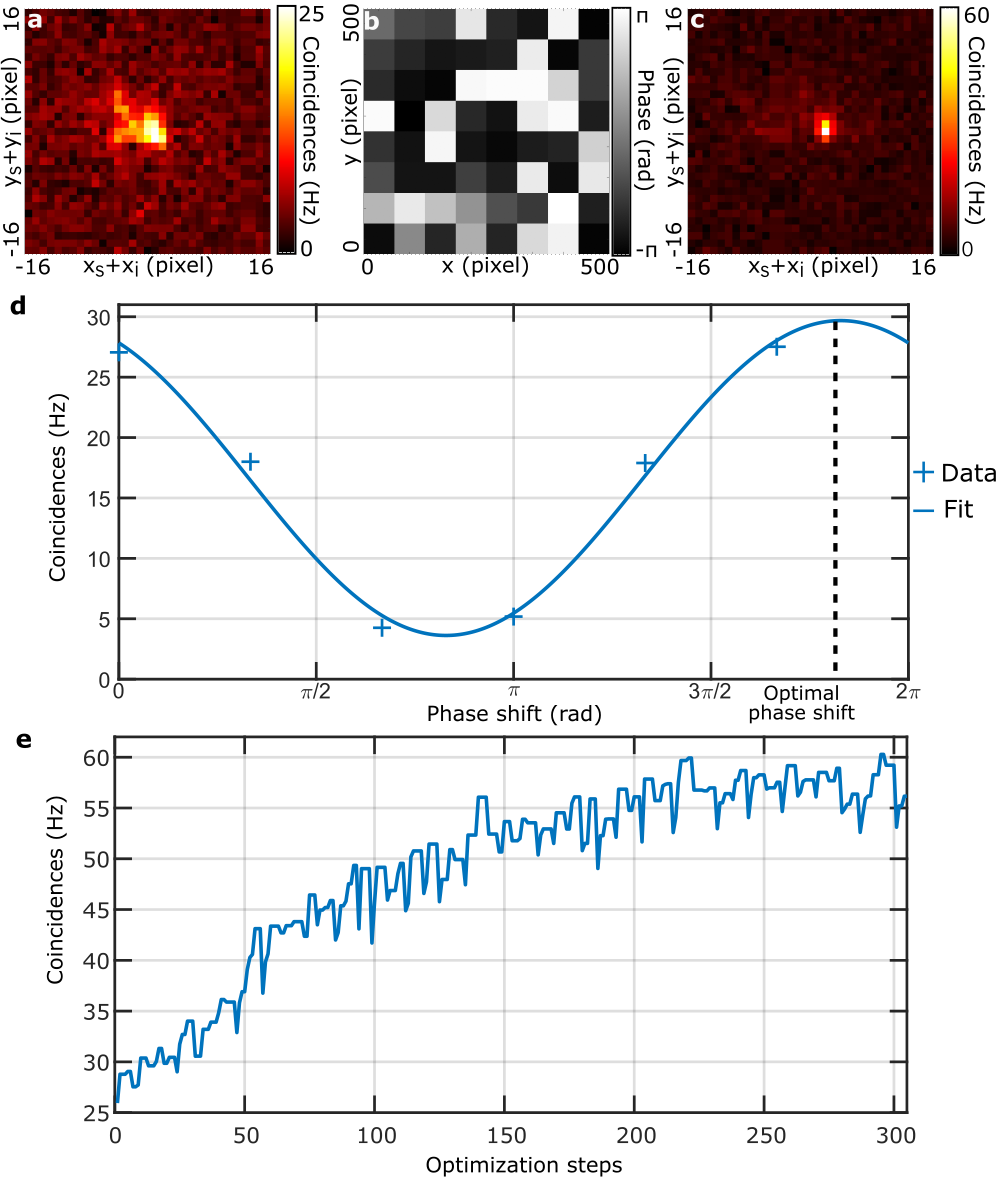}
    \caption{\textbf{Experimental results for the single-scattering regime.} \textbf{(a)} Sum-coordinate projection of $G^{(2)}$ acquired in the presence of the Parafilm layer with a flat phase applied to the SLM. \textbf{(b)} Final optimized phase mask. \textbf{(c)} Sum-coordinate projection of $G^{(2)}$ acquired with the optimized phase mask displayed on the SLM, demonstrating a partial restoration of spatial correlations between the photon pairs. \textbf{(d)} Central correlation value $C_0$ as a function of the phase shift $ \Delta\theta$ applied to a randomly selected subset of macropixels, along with the corresponding cosine fit. \textbf{(e)} Optimization curve showing the evolution of $C_0$ as a function of the number of optimization steps. For optimization, each $G^{(2)}$ measurement required $3.75.10^6$ frames at $7~\mu$s exposure.}
    \label{Figure3}
\end{figure}

\textit{Single-scattering regime ---} To investigate the single-scattering regime, we replace the PDMS with a Parafilm layer of thickness $L$ comparable to its scattering mean free path ($l_s\sim100\mu$m $\sim L$)~\cite{boniface_noninvasive_2019}. This degrades the peak in the $G^{(2)}$ sum-coordinate projection (Fig.~\ref{Figure3}a). 
To restore the spatial correlations, we use a random partitioning algorithm~\cite{vellekoop_focusing_2007}, with the correlation value $C_0$ as the optimization metric and the SLM divided into $8\times8$ active macropixels. 
This optimization yields a final phase mask (Fig.~\ref{Figure3}b) that successfully restores the correlation peak (Fig.~\ref{Figure3}c). 
The optimal mask is obtained iteratively: at each step, a random subset of half the macropixels is sequentially modulated over $7$ phase shifts, and the corresponding $C_0$ values are fitted with a cosine function to determine the phase maximizing the correlation (Fig.~\ref{Figure3}d). 
This procedure is repeated with a new random subset at each iteration, reaching convergence after 300 steps (Fig.~\ref{Figure3}e). 
In the scattering regime, the correction efficiency is quantified through the enhancement ratio $\eta$, defined as the ratio of the optimized to initial $C_0$ values~\cite{shekel_fundamental_2025}. 
We obtain $\eta \approx 2.4$. More details about this experiment are provided in Section~IV of the Supplemental Material.

\textit{Multiple-scattering regime ---} The multiple-scattering regime is investigated using numerical simulations. For this, we use an experimentally measured transmission matrix of a thick scattering medium (6 Parafilm layers stuck together, $L \sim 6 l_s$) connecting $16 \times 16$ SLM input modes to $32 \times 32$ camera pixels~\cite{popoff_measuring_2010}. 
The quantum experiment is simulated via the matrix formalism~\cite{courme_manipulation_2023}:
\begin{equation}
    \Psi_{\text{out}} = \mathcal{F} D_{\text{SLM}} \Psi_{\text{in}} T_m^t,
\end{equation}
where $\Psi_{\text{in}}$ and $\Psi_{\text{out}}$ are the input and output two-photon wave-function matrices, $D_{\text{SLM}}$ is the SLM diagonal phase matrix and $T_m$ is the measured transmission matrix. $T_m$ encompasses $T$, that is the matrix linking the input to output surface of the scattering medium, and the subsequent signal-arm Fourier transform. 
$\mathcal{F}$ denotes the optical Fourier transform in the idler arm.

\begin{figure}[h!]
    \centering
    \includegraphics[width=1\linewidth]{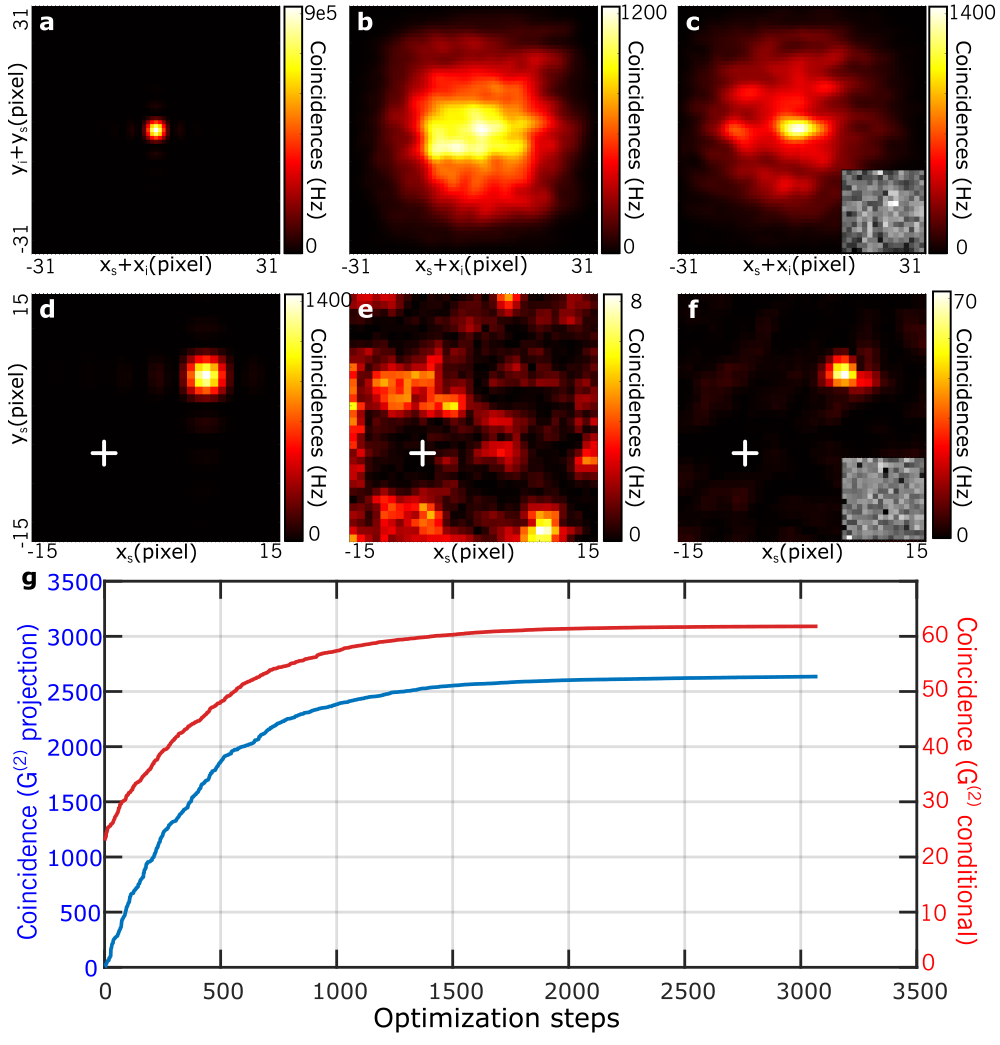}
    \caption{\textbf{Simulation results for the multiple-scattering regime.} Sum-coordinate projection of $G^{(2)}$ simulated \textbf{(a)} without the scattering medium, \textbf{(b)} with the scattering medium prior to optimization, and \textbf{(c)} after optimization. The optimal phase mask is shown in inset. Conditional correlation images $G^{(2)}(\vec{r}_s,-\vec{r}_0)$ for $\vec{r}_0=(-5,-7)$, simulated \textbf{(d)} without the scattering medium, \textbf{(e)} with the scattering medium prior to optimization, and \textbf{(f)} after optimization. The position of $\vec{r}_0$ is marked with a white cross and the corresponding optimal phase mask is shown in inset. \textbf{(g)} Optimization curves for focusing on the sum-coordinate projection of $G^{(2)}$ (blue) and on the conditional projection (red). The corresponding vertical axes are shown on the left and right sides of the graph.}
    \label{Figure4}
\end{figure}

Optimization is performed via a random partitioning algorithm, using two distinct feedback metrics. 
First, we use the central correlation value $C_0$, yielding a weakly restored peak with an enhancement of $\eta = 1.7$. The corresponding $G^{(2)}$ sum-coordinate projections -- measured without the medium, before optimization, and after optimization -- are shown in Figs.~\ref{Figure4}a-c, respectively. 
Alternatively, we use the correlation of a single pair of symmetric pixels $(\vec{r}_s,\vec{r}_i)=(\vec{r}_0,-\vec{r}_0)$ as the feedback metric, where $\vec{r}_0$ is an arbitrary position. 
It efficiently restores the correlation, yielding $\eta=49.5$. 
The corresponding conditional correlation images $G^{(2)}(\vec{r}_s,-\vec{r}_0)$ -- representing the probability of detecting a signal photon at $\vec{r}_s$ given an idler detection at $-\vec{r}_0$ -- are shown without the medium, before, and after optimization in Figs.~\ref{Figure4}d-f, respectively. 
The optimization dynamics for both approaches are compared in Fig.~\ref{Figure4}g, showing the convergence curves when optimizing $C_0$ (blue) and the targeted single pair (red). 
In the latter case, note that the optimization remains effective regardless of the specific target pixel pair $(\vec{r}_s,\vec{r}_i)$ selected (even for asymmetric pixels).

\textit{Theory ---} In the low-order aberration regime, a Strehl ratio of $1$ is theoretically achievable provided the SLM (or DM) has sufficient control degrees of freedom (DoFs) to fully span the aberration modes of the system. 
In the scattering regime, however, perfect correction is generally not achievable because the control DoFs (e.g. SLM macropixels) are fewer than the number of spatial modes supported by the medium, and because they only shape the phase of the field.
To evaluate performance of our optimization approach, we define the theoretical enhancement factor $\eta_t$ as the ratio of the disorder-averaged metric $C_0$ after and before optimization~\cite{shekel_fundamental_2025}. 

In the single-scattering regime, the transmission matrix is modeled as a diagonal matrix ($T_{kl} = |T_{kl}| e^{i \phi_{kl}} \delta_{kl}$), with elements $T_{kl}$ drawn from independent and identically distributed circular complex Gaussian distributions. 
Assuming uniform signal photon intensity across the SLM and near-perfect spatial correlations at the medium input ($\psi(\vec{r}_s,\vec{r}_i) = \delta(\vec{r}_s-\vec{r}_i)$), the expected enhancement is:
\begin{equation}
\label{Eqenhnacement1}
\eta_t = 1 + \frac{\pi}{4} (N-1),
\end{equation}
where $N$ is the number of active SLM macropixels. Interestingly, this result is identical to that of classical wavefront shaping~\cite{vellekoop_focusing_2007}. Full derivation of Eq.~\eqref{Eqenhnacement1} is provided in Section~II of the Supplemental Material).

Experimentally (Fig.~\ref{Figure3}), the measured enhancement ($\eta \approx 2.4$) is approximately one order of magnitude lower than predicted ($\eta_t \approx 50$ for $N=64$). 
This discrepancy primarily stems from $G^{(2)}$ projection noise that strongly limits the algorithm convergence and the challenge of maintaining long-term sample stability over the $300$ optimization steps (12 days of acquisition). 
While these measurements demonstrate experimental feasibility, they do not quantitatively validate the theoretical model; however, numerical simulations in Section~III of the Supplemental Material confirm Eq.~\eqref{Eqenhnacement1}.

In the multiple-scattering regime, $T$ is non-diagonal ($T_{kl} = |T_{kl}| e^{i \phi_{kl}}$), and the enhancement factor becomes:
\begin{equation}
\label{Eqenhnacement2}
\eta_t = 1 + \frac{\pi(N-1)}{4M^b},
\end{equation}
where $b \approx 0.83$ and $M$ is the number of independent output spatial modes i.e. camera pixels in the region of interest when the imaging magnification is properly adjusted. 
In the simulation shown in Figs.~\ref{Figure4}a-c, the measured enhancement ($\eta=1.7$) matches well with theory ($\eta_t=1.63$ for $N=256$ input modes and $M=1024$ output modes). 
The analytical derivations and complementary simulations supporting Eq.~\eqref{Eqenhnacement2} are provided in Sections~II and III of the Supplemental Material.

Finally, we note that Equation~\eqref{Eqenhnacement2} predicts a limited enhancement compared to the single-scattering case, as it is reduced by a factor of $M^b>>1$. 
Indeed, recovering a large value for $C_0$ - which corresponds to the sum of all anti-diagonal elements of $\Psi_{\mathrm{out}}$ - would require increasing all the correlation values along the anti-diagonal; in other words, it would require simultaneously refocusing the correlations between all anti-symmetric mode pairs on the camera. 
However, in the multiple-scattering regime, each entangled mode pair generates a distinct two-photon speckle pattern. 
Achieving this would therefore require refocusing many different speckles using a single phase pattern, which exceeds the control DoFs of a single SLM. 
Consequently, the optimization process converges towards a solution in which only a few of these speckles are partially optimized, leaving the $G^{(2)}$ peak predominantly as an incoherent sum of speckles with minimal overall enhancement.
By contrast, restricting the optimization to a single mode pair (Figs.~\ref{Figure4}d-f) isolates the process, yielding a much larger enhancement consistent with Eq.~\eqref{Eqenhnacement1}. More details are provided in Section~II of the Supplemental Material.

\textit{Conclusion ---} We have experimentally and numerically demonstrated nonlocal wavefront shaping using spatially entangled photon pairs. 
By modulating the wavefront of one photon while its entangled partner propagates through a complex medium, we restore their spatial correlations across low-order aberrations, single scattering, and multiple scattering. 
We developed an analytical model predicting the maximum achievable enhancement in each regime, which we used to interpret our results.

Currently, the practical implementation of our method is limited by measurement noise, requiring simultaneous improvements in detection sensitivity, acquisition speed, and source brightness. 
A promising route is the use of emerging single-photon time-stamping cameras, such as the Tpxcam series~\cite{nomerotski_imaging_2019,hogenbirk_intensified_2026}, which could substantially improve the signal-to-noise ratio of each correlation measurement. 
Combined with advanced signal-processing and machine-learning approaches~\cite{lotan_sparsity-driven_2025,xu_learning_2026}, as well as optimized wavefront-shaping algorithms~\cite{conkey_genetic_2012}, these technologies could enable efficient nonlocal optimization.

The primary advantage of our approach is the physical decoupling of the imaging path from the aberration-correction apparatus. 
This spatial separation could be particularly beneficial for imaging through scattering environments where integrating a wavefront-correction system into the primary optical setup is unfeasible or challenging. 
Such scenarios include inherently compact industrial imaging systems, such as microscopes, or situations requiring correction architectures significantly more complex than a single SLM, such as multi-plane configurations~\cite{a_rocha_self-configuring_2025}. 
Ultimately, these nonlocal control techniques open new avenues for advanced optical imaging and sensing in complex media.

\vspace{\baselineskip}

\noindent \textbf{Acknowledgments}
H.D. acknowledges funding from the ERC Starting Grant (No. SQIMIC-101039375) and the ANR (No. ANR-24-CE97-0001 and ANR-23-CE47-0014).
A.F. acknowledges SA QuTI, CSIR Rental Pool and OMT for financial support.\\
\noindent \textbf{Author Contributions}
Y.T. analyzed the data, designed and performed the experiments, with support from P.O, N.G and P.C. H.D. and A.F. conceived the original idea. All authors discussed the results and contributed to the manuscript. H.D supervised the project.

\clearpage
\onecolumngrid
\begin{center}
  \textbf{\large Supplemental Material: Nonlocal wavefront shaping through complex media}\\[.2cm]
  Yanis Trouyet$^1$, Neelan Gounden$^2$, Pedro Ornelas$^2$, Patrick Cameron$^{3,4}$, Andrew Forbes$^2$, and Hugo Defienne$^1$\\[.1cm]
  {\itshape $^1$Sorbonne Université, CNRS, Institut des NanoSciences de Paris, INSP, F-75005 Paris, France\\
  $^2$University of the Witwatersrand, 9 Enoch Sontonga Ave, Johannesburg, South Africa\\
  $^3$Dipartimento di Fisica, Università degli Studi di Napoli Federico II, Complesso Universitario di Monte Sant’Angelo, Via Cintia, 80126 Napoli, Italy\\
  $^4$Scuola Superiore Meridionale, Via Mezzocannone, 4, 80138 Napoli, Italy\\}
\end{center}
\vspace{1cm}

\setcounter{equation}{0}
\setcounter{figure}{0}
\setcounter{table}{0}
\setcounter{page}{1}
\renewcommand{\theequation}{B\arabic{equation}}
\renewcommand{\thefigure}{S\arabic{figure}}
\renewcommand{\thesection}{\Roman{section}}
\renewcommand{\thesubsection}{\Alph{subsection}}
\renewcommand{\thesubsubsection}{\roman{subsubsection}}

\section{$G^{(2)}$ measurement and sum-coordinate projection}

\subsection{Spatial $G^{(2)}$ measurement} 

Detection of the photon pairs is performed using a single-photon avalanche diode (SPAD) camera. 
During an acquisition sequence, the camera records a set of $P$ frames, denoted $I_p$ ($p \in [1,P]$). 
In each frame, every pixel yields a binary value ($0$ or $1$) indicating whether a photon detection event was registered. 
To evaluate the spatial correlation function $G^{(2)}(\vec{r}_s,\vec{r}_i)$, where $\vec{r}_s$ and $\vec{r}_i$ denote the transverse positions of two distinct camera pixels, we use the following expression:
\begin{equation}
    G^{(2)}(\vec{r_s},\vec{r_i}) = \frac{1}{P}\sum_{p=1}^P I_p(\vec{r_s}) I_p(\vec{r_i}) - \frac{1}{P^2}\left ( \sum_{p=1}^P I_p(\vec{r_s}) \right) \left ( \sum_{p=1}^P I_p(\vec{r_i}) \right).
\end{equation}
Here, the first term represents the frame-by-frame product of the pixel values summed over the entire acquisition set, which estimates the total measured coincidences. From this, we subtract a second term - the product of the total intensities at each pixel - to account for accidental coincidences. 
Further details regarding the measurement of $G^{(2)}$ with this type of SPAD camera can be found in Ref.~\cite{ndagano_imaging_2020}.

\subsection{$G^{(2)}$ sum-coordinate projection} 

The correlation images shown in the manuscript, from which the metric $C_0$ is extracted, are projections of $G^{(2)}$ along the sum-coordinate axis. These will be denoted as $C(\vec{r}_+)$. 

In theory, this projection is defined as follows~\cite{defienne_general_2018}:
\begin{equation}
\label{eqSM1}
    C(\vec{r}_+) = \iint G^{(2)}(\vec{r}, \vec{r}_+ - \vec{r}) \, d\vec{r}.
\end{equation}
To better understand it, we perform the following change of variables: $\vec{r}_+ = (\vec{r}_s + \vec{r}_i)/2$ and $\vec{r}_- = (\vec{r}_s - \vec{r}_i)/2$. Then, we define the function $G_\pm^{(2)}$ such that $G_\pm^{(2)}(\vec{r}_+, \vec{r}_-) = G^{(2)}(\vec{r}_i, \vec{r}_s)$. With this definition, Equation~\eqref{eqSM1} can be rewritten as:
\begin{equation}
    C(\vec{r}_+) = \iint G^{(2)}(\vec{r}, \vec{r}_+ - \vec{r}) \, d\vec{r} = \iint G_\pm^{(2)}(\vec{r}_+, \vec{r}_-) \, d\vec{r}_-.
\end{equation}
From the above equation, we understand that the sum-coordinate projection $C$ is obtained by integrating the second-order correlation function along the diagonal $\vec{r}_- = \vec{r}_s - \vec{r}_i$, thereby projecting it onto the sum-coordinate $\vec{r}_+ = \vec{r}_s + \vec{r}_i$. 
Physically, the value $C(\vec{r}_+)$ can be interpreted as the average probability of detecting coincident photons across all camera pixel pairs that are symmetric with respect to the coordinate $\vec{r}_+$. In particular, if the photon pairs arrive at the camera in an anti-correlated manner, as is the case in a Fourier configuration without aberrations (Fig.~1 of the manuscript), a sharp peak is observed at the center of the sum-coordinate projection (Fig.~1d of the manuscript).

In practice, the function $G^{(2)}$ is measured using a camera and is therefore discrete. Moreover, its integration space is finite. Consequently, it is given by: 
\begin{equation}
\label{eqSM2}
    C(\vec{r}_+) = \sum_{x=1}^{\sqrt{M}} \sum_{y=1}^{\sqrt{M}} G^{(2)}(x, y, x_+ - x, y_+ - y),
\end{equation}
where $\vec{r}=(x,y)$ and $\vec{r}_+ = (x_+,y_+)$ denote the transverse spatial coordinates of the camera pixels, and $M$ is the total number of pixels in the region of interest of the camera where the measurement is performed. 

\section{Analytical derivation of the enhancement ratio}

This section provides an analytical demonstration of Equations (1), (2) and (3) of the main text. 

\subsection{Demonstration of Equation (1) of the main text}

In the low-gain regime and under the paraxial approximation, the field state at the crystal output, $|\psi_{in}\rangle$, can be written in the following general form~\cite{walborn_spatial_2010}:
\begin{equation}
   |\psi_{in}\rangle = \iint d\vec{r}_i d\vec{r}_s \psi_{in}(\vec{r}_i,\vec{r}_s) a_s^\dagger(\vec{r}_s) a_i^\dagger(\vec{r}_i) |0\rangle, 
\end{equation}
where $\psi_{in}(\vec{r}_i,\vec{r}_s)$ is the biphoton amplitude at the output surface of the crystal for transverse positions $\vec{r}_i$ and $\vec{r}_s$, and $a_{s}^\dagger(\vec{r}_s)$ and $a_{i}^\dagger(\vec{r}_i)$ are the signal and idler photon creation operators at those respective transverse coordinates. 
For clarity, the vacuum term and the normalization constant have been omitted from the expression for $|\psi_{in}\rangle$.

After propagation through the optical setup depicted in Fig.~1, the biphoton amplitude $\psi_{\text{out}}(\vec{r}'_i,\vec{r}'_s)$ at the camera plane is given by~\cite{abouraddy_entangled-photon_2002}:
\begin{equation}
\label{propagation}
    \psi_{\text{out}}(\vec{r}'_i,\vec{r}'_s) = \iint d\vec{r}_i d\vec{r}_s h_i(\vec{r}'_i,\vec{r}_i) h_s(\vec{r}'_s,\vec{r}_s) \psi_{\text{in}}(\vec{r}_i,\vec{r}_s),
\end{equation}
where $h_i$ and $h_s$ denote the coherent point spread functions (PSFs) of the idler and signal arms, respectively, linking the output plane of the crystal to the camera plane. 
By discretizing this system, the continuous transformations in Equation~\eqref{propagation} can be expressed through a matrix formalism~\cite{popoff_measuring_2010,courme_manipulation_2023}. The continuous PSFs along the signal and idler arms can be modeled by the following matrix products:
\begin{align}
h_s &\rightarrow \mathcal{F}_5 \mathcal{F}_4 \mathcal{F}_3 T \mathcal{F}_2 \mathcal{F}_1, \\
h_i &\rightarrow \mathcal{F}_8 \mathcal{F}_7 \mathcal{F}_6 D_{\text{SLM}} \mathcal{F}_2 \mathcal{F}_1,
\end{align}
where $\mathcal{F}_p$ represents the matrix associated with the $p$-th $2f$-lens system performing a Fourier transform, $D_{\text{SLM}}$ is the matrix associated with the SLM, and $T$ is the transmission matrix of the scattering medium connecting its input plane to its output plane. As a result, the biphoton amplitude can be represented by the matrix $\Psi_{in}$ and $\Psi_{out}$, linked as follows: 
\begin{equation}
\label{matrixfull}
    \Psi_{out} = \mathcal{F}_8 \mathcal{F}_7 \mathcal{F}_6 D_{\text{SLM}} \mathcal{F}_2 \mathcal{F}_1 \Psi_{in} (\mathcal{F}_5 \mathcal{F}_4 \mathcal{F}_3 T \mathcal{F}_2 \mathcal{F}_1)^t. 
\end{equation}
Equation~(1) of the main text is obtained by simplifying Eq.~(\ref{matrixfull}). Specifically, we neglect the relay telescopes and their associated magnification effects (i.e. omitting $\mathcal{F}_2 \mathcal{F}_1$, $\mathcal{F}_7 \mathcal{F}_6$, and $\mathcal{F}_4 \mathcal{F}_3$) and set $\mathcal{F}_8 = \mathcal{F}_5 = \mathcal{F}$. In the manuscript, the product $\mathcal{F}T$ is then replaced directly by $T_m$, where $T_m$ is an experimentally measured transmission matrix encompassing both the scattering medium and the subsequent lens.

\subsection{Demonstration of Equations (2) and (3) of the main text}

For simplicity and without loss of generality, we denote each spatial mode by a single index $l$, without explicitly distinguishing between the two spatial dimensions.
Expanding upon Eq.~(1) of the main text, the two-photon correlation function $G^{(2)}_{ll'}$ between two transverse spatial modes $l$ and $l'$ can be expressed as:
\begin{equation}
\label{eqqq1}
    G^{(2)}_{ll'} = \left| \sum_{k,k',k''} f_{lk} e^{i \theta_k} \Psi_{kk'} T_{k''k'} f_{l'k''} \right|^2,
\end{equation}
where $f_{lk} = \exp[-2 \pi i (\vec{r}_k \cdot \vec{r}_l)/(\lambda f)]$ denotes the elements of the Fourier matrix $\mathcal{F}$ linking spatial modes $l$ and $k$ (associated with transverse coordinates $\vec{r}_l$ and $\vec{r}_k$, where $f$ is the focal length of the corresponding lens).
Additionally, $T_{k'l'}$ and $\Psi_{kk'}$ are the matrix elements of the transmission matrix $T$ and the input two-photon state $\Psi_{\text{in}}$, respectively, and $\theta_k$ is the phase imparted by the SLM on transverse spatial mode $k$. Finally, the discrete version of the sum-coordinate projection of $G^{(2)}$, noted $C^+_p$, can be written as:
\begin{equation}
\label{fullqr}
   C^+_p = \sum_{l} G^{(2)}_{l \, p-l} =  \sum_l \left| \sum_{k,k',k''} f_{lk} e^{i \theta_k} \Psi_{kk'} T_{k''k'} f_{p-l \, k''} \right|^2.
\end{equation}

For the rest of the derivations, we assume perfect spatial correlations between photons at the SLM plane i.e. $\Psi_{kl} = \delta_{kl}$~\cite{abouraddy_entangled-photon_2002}. Experimentally, this implies that each active SLM macropixel used for optimization is significantly larger than the transverse spatial correlation width of the photon pairs. 
Under these conditions, the system dimensionality $N$ is defined entirely by the number of SLM macropixels, and Eqs.~\eqref{eqqq1} and \eqref{fullqr} simplify to:
\begin{equation}
    G^{(2)}_{ll'} = \left| \sum_{k,k'}^N f_{lk} e^{i \theta_k} T_{k'k} f_{l'k'} \right|^2
\end{equation}
and
\begin{equation}
\label{fullq}
   C^+_p = \sum_{l=1}^M G^{(2)}_{l \, p-l} =  \sum_l \left| \sum_{k,k'}^N f_{lk} e^{i \theta_k} T_{k'k} f_{p-l \, k'} \right|^2,
\end{equation}
where $M$ denotes the number of spatial modes included in the summation used to compute the sum-coordinate projection of $G^{(2)}$. 
Experimentally, $M$ corresponds to the number of camera pixels defining the detection region over which photon pairs are detected to measure $G^{(2)}$.

\subsubsection{Demonstration of Equation (2) of the manuscript} 

Here, we consider the single-scattering regime. To do so, we rely on the following assumptions:
\begin{enumerate}
\item The scattering medium is modeled by a diagonal transmission matrix: $T_{kl} = |T_k| e^{i \phi_k} \delta_{kl}$.
\item The amplitude $|T_k|$ and phase $\phi_k$ coefficients are independent and identically distributed (i.i.d.) random variables drawn from a circular complex Gaussian distribution~\cite{goodman_speckle_2007}.
\end{enumerate}
Equation~\eqref{fullq} then simplifies to:
\begin{equation}
C^+_p = \sum_{l=1}^M \left| \sum_{k=1}^N f_{lk} f_{p-l , k} |T_k| e^{i (\phi_k+\theta_k)} \right|^2.
\end{equation}
Based on our definition of the enhancement factor, we first calculate the disorder-averaged correlation value at position $p=0$ i.e. $\langle C^+_0 \rangle$:
\begin{align}
\langle C^+_0 \rangle &= \sum_{l=1}^M \left[ \sum_{k=1}^N \langle |T_k|^2 \rangle + \sum_{k \neq k'}^N f_{lk} f_{-l , k} f_{lk'}^* f_{-l , k'}^* e^{i (\theta_k-\theta_{k'})} \langle |T_k| e^{i \phi_k} \rangle \langle |T_{k'}| e^{-i \phi_{k'}} \rangle \right] \nonumber \\ 
&= \sum_{l=1}^M \left[ \sum_{k=1}^N \langle |T|^2 \rangle \right] \nonumber \\ 
&= M N \langle |T|^2 \rangle,
\end{align}
where $\langle |T|^2 \rangle$ is defined as the variance of the amplitude element of $T$. Here, we used the property that the transmission matrix coefficients are statistically independent and satisfy $\langle |T_k| e^{i\phi_k} \rangle = 0$.

Second, we evaluate the disorder-averaged optimized correlation value $\langle C_0^+\rangle_{\mathrm{opt}}$. 
In the single-scattering regime, the optimal SLM phase pattern exactly compensates for the phase introduced by the scattering medium i.e. $\theta_k = -\phi_k$. This yields the following expression:
\begin{align}
\langle C^+_0 \rangle_{opt} &= \sum_{l=1}^M \left[ \sum_{k=1}^N \langle |T_k|^2 \rangle + \sum_{k \neq k'}^N f_{lk} f_{-l , k} f_{lk'}^* f_{-l , k'}^* \langle |T_k| \rangle^2 \right] \nonumber \\ 
&= \sum_{l=1}^M \left[ N\langle |T|^2 \rangle  + N(N-1) \langle |T| \rangle^2 \right] \nonumber \\ 
&= M [ N \langle |T|^2 \rangle + N(N-1)  \langle |T| \rangle^2 ],
\end{align}
where we used the following result:
\begin{align}
 \sum_{k \neq k'}^N f_{lk} f_{-l , k} f_{lk'}^* f_{-l , k'}^* &=  \left| \sum_{k=1}^N f_{lk} f_{-l , k} \right|^2 -  \sum_{k=1}^N \left | f_{lk} f_{-l , k} \right|^2   \nonumber \\ 
&= \left| \sum_{k=1}^N f_{lk} f_{ k,l}^* \right|^2 -  N \nonumber \\ 
&= N^2 -  N \nonumber \\.
\end{align}
Finally, we obtain the enhancement ratio in the single-scattering regime:
\begin{equation}
    \eta_t = \frac{\langle C^+_0 \rangle_{opt}}{\langle C^+_0 \rangle} = 1 + \frac{\pi}{4}{(N-1)},
\end{equation}
where we used the fact that $\frac{\langle |T| \rangle^2}{\langle |T|^2 \rangle} = \frac{\pi}{4}$ for complex gaussian variables i.i.d.

\subsubsection{Demonstration of Equation (3) of the manuscript} 
\label{equation3section}

In the multiple-scattering case, we consider that the scattering medium is described by a transmission matrix $T_{kl} = |T_{kl}| e^{i \phi_{kl}}$ where its elements are complex circular Gaussian random variable.
Equation~\eqref{fullq} then simplifies to:
\begin{equation}
\label{fullqqq}
C^+_p = \sum_{l=1}^M \left| \sum_{k=1}^N  e^{i \theta_k} \sum_{k'=1}^N f_{lk} T_{k'k} f_{p-l \, k'} \right|^2 = \sum_{l=1}^M \left| \sum_{k=1}^N e^{i \theta_k} \alpha_{kpl} \right|^2,
\end{equation}
where
\begin{equation}
\alpha_{kpl} = \sum_{k'=1}^{N} f_{lk} T_{k'k} f_{p-l,\,k'}
\end{equation}
is also a complex random variable with statistical properties similar to those of $T_{kl}$, in particular $\langle \alpha_{kpl} \rangle = 0$, $\langle |\alpha_{kpl}|^2 \rangle = \langle |T|^2 \rangle$, and $\langle |\alpha_{kpl}| \rangle^2 = \langle |T| \rangle^2 = (\pi/4)\langle |T|^2 \rangle$. 

We first calculate the disorder-averaged correlation value at position $p=0$ i.e. $\langle C^+_0 \rangle$:
\begin{align}
\label{fullq22}
\langle C^+_0 \rangle &= \sum_{l=1}^M \left[ \sum_{k=1}^N \langle |\alpha_{k,-l,l}|^2 \rangle + \sum_{k \neq k'}^N e^{i (\theta_k-\theta_{k'})} \langle \alpha_{k,-l,l} \rangle \langle \alpha_{k',-l,l}^* \rangle \right] \nonumber \\ 
&= \sum_{l=1}^M \left[ \sum_{k=1}^N \langle |T|^2 \rangle \right] \nonumber \\ 
&= M N \langle |T|^2 \rangle.
\end{align}

Next, we evaluate the disorder-averaged optimized correlation value $\langle C_0^+\rangle_{\mathrm{opt}}$. As can be seen from Eq.~\eqref{fullqqq}, the SLM optimizes a sum of $M$ terms, each given by the squared modulus of a sum of $N$ random phasors. Since these terms are statistically independent, this optimization process relates to problems found in classical wavefront shaping for multi-target optimization~\cite{vellekoop_phase_2008} or total light transmission control~\cite{popoff_coherent_2014}.

To calculate the theoretical enhancement, we initially assume that the optimal SLM phase pattern correspond to the phase pattern that maximizes the contribution of the single term with the largest amplitude. Assuming that this dominant term corresponds to the index $l=l_0$, we obtain:
\begin{align}
\label{fullq222}
\langle C^+_0 \rangle_{\mathrm{opt}} &= \left| \sum_{k=1}^N |\alpha_{kpl}| \right|^2 + \sum_{l\neq l_0}^M \left| \sum_{k=1}^N \alpha_{kpl} \right|^2 \nonumber \\
&= N \langle |T|^2 \rangle + N(N-1)  \langle |T| \rangle^2 + (M-1) N \langle |T|^2 \rangle \nonumber \\
&= N(N-1)  \langle |T| \rangle^2 + M N \langle |T|^2 \rangle.
\end{align}
Finally, we obtain the expected enhancement ratio in the multiple-scattering regime:
\begin{equation}
\label{formulahp}
    \eta_t = \frac{\langle C^+_0 \rangle_{\mathrm{opt}}}{\langle C^+_0 \rangle} = 1 + \frac{\pi}{4}\frac{(N-1)}{M}.
\end{equation}

However, the numerical simulations performed in Section~\ref{simulationsthick} reveal that while the linear dependence on $N-1$ is perfectly preserved, the slope deviates from $\frac{\pi}{4M}$ for $M>1$. Empirically, we find that the enhancement factor is instead well described by an equation of the form:
\begin{equation}
    \eta_t = \frac{\langle C^+_0 \rangle_{\mathrm{opt}}}{\langle C^+_0 \rangle} = 1 + \frac{\pi}{4}\frac{(N-1)}{M^b},
\end{equation}
where $b \approx 0.83<1$. The value of $b$ is obtained by fitting the simulations, as shown in Fig.~\ref{FigureSM2}c. This discrepancy arises because the aforementioned assumption - that the optimal SLM pattern found by optimization corresponds to the one that solely maximizes the single largest amplitude term - is false. More details are provided in Section~\ref{simulationsthick}.

\subsubsection{Optimization on a single pair of spatial modes in the multiple scattering regime}

Finally, we consider the specific optimization case studied in Figs.~4d-f of the manuscript. 
Instead of choosing $C_0$ as the feedback metric, we directly consider the value $G^{(2)}_{lp}$ between a pair of arbitrarily chosen spatial modes $l$ and $p$. 
This metric is denoted as $G^{(2)}_0$:
\begin{equation}
\label{eqB24}
G^{(2)}_0 = G^{(2)}_{lp} = \left| \sum_{k=1}^N  e^{i \theta_k} \sum_{k'=1}^N f_{lk} T_{k'k} f_{p , k'} \right|^2 = \left| \sum_{k=1}^N e^{i \theta_k} \beta_{klp} \right|^2,
\end{equation}
where 
\begin{equation}
\beta_{klp} = \sum_{k'=1}^{N} f_{lk} T_{k'k} f_{p,,k'}
\end{equation}
is also a complex random variable with statistical properties similar to those of $T_{kl}$; 
in particular, $\langle \beta_{klp} \rangle = 0$, $\langle \vert{}\beta_{klp}\vert{}^2 \rangle = \langle \vert{}T\vert{}^2 \rangle$, and $\langle \vert{}\beta_{klp}\vert{} \rangle^2 = \langle \vert{}T\vert{} \rangle^2 = (\pi/4)\langle \vert{}T\vert{}^2 \rangle$.

Using the same mathematical approach as in previous paragraphs, we first calculate $\langle G^{(2)}_0 \rangle$ in the multiple scattering regime, which can be easily deduced from Eq.~\eqref{fullq22}:
\begin{equation}\langle G^{(2)}_0 \rangle = N \langle |T|^2 \rangle.
\end{equation}

Then, we evaluate the disorder-averaged optimized correlation value $\langle G^{(2)}_0 \rangle_{opt}$. 
In this case, the optimal SLM phase pattern is the one that aligns all the phasor terms $\beta_{klp}$ i.e. that cancels their respective phase terms. 
Thus, we have:
\begin{equation}
\langle G^{(2)}_0 \rangle_{opt} = \left| \sum_{k=1}^N |\beta_{klp}| \right|^2 = N \langle |T|^2 \rangle + N(N-1) \langle |T| \rangle^2,
\end{equation}
a result deduced from the calculations already performed in Eq.~\eqref{fullq222}. 

Finally, we obtain the enhancement ratio in the form:
\begin{equation}
\eta_t = \frac{\langle G^{(2)}_0 \rangle_{opt}}{\langle G^{(2)}_0 \rangle} = 1 + \frac{\pi}{4}(N-1).
\end{equation}

\section{Additional simulations}

The complementary simulations presented in this section aim to verify Eqs.~(2) and (3) of the manuscript, in particular the dependence of the enhancement factor on $N$ and $M$. 

\subsection{Simulations in the single-scattering regime}

To model a thin scattering medium, we first generate a transmission matrix whose complex elements are independent and identically distributed (i.i.d.) random variables drawn from a circular complex Gaussian distribution. In practice, this is generated using the Matlab command \texttt{(randn(Ns, Ns) + 1i * randn(Ns, Ns)) / sqrt(2)}, where $N_s=1000$ is the dimension of the simulation space. 
Then, we retain only the diagonal elements of this matrix, setting all off-diagonal terms to zero. 
The propagation of the two-photon state through this scattering medium is then computed using Eq.~(1) of the main manuscript. We initialize the system with a diagonal input matrix $\Psi_{\text{in}}$ to simulate perfect spatial correlations, and use a discrete Fourier transform matrix for $\mathcal{F}$. The SLM is modeled by a diagonal matrix whose elements encode the dynamically reconfigurable phase shifts. 

\begin{figure}[h!]
    \centering
    \includegraphics[width=0.7\linewidth]{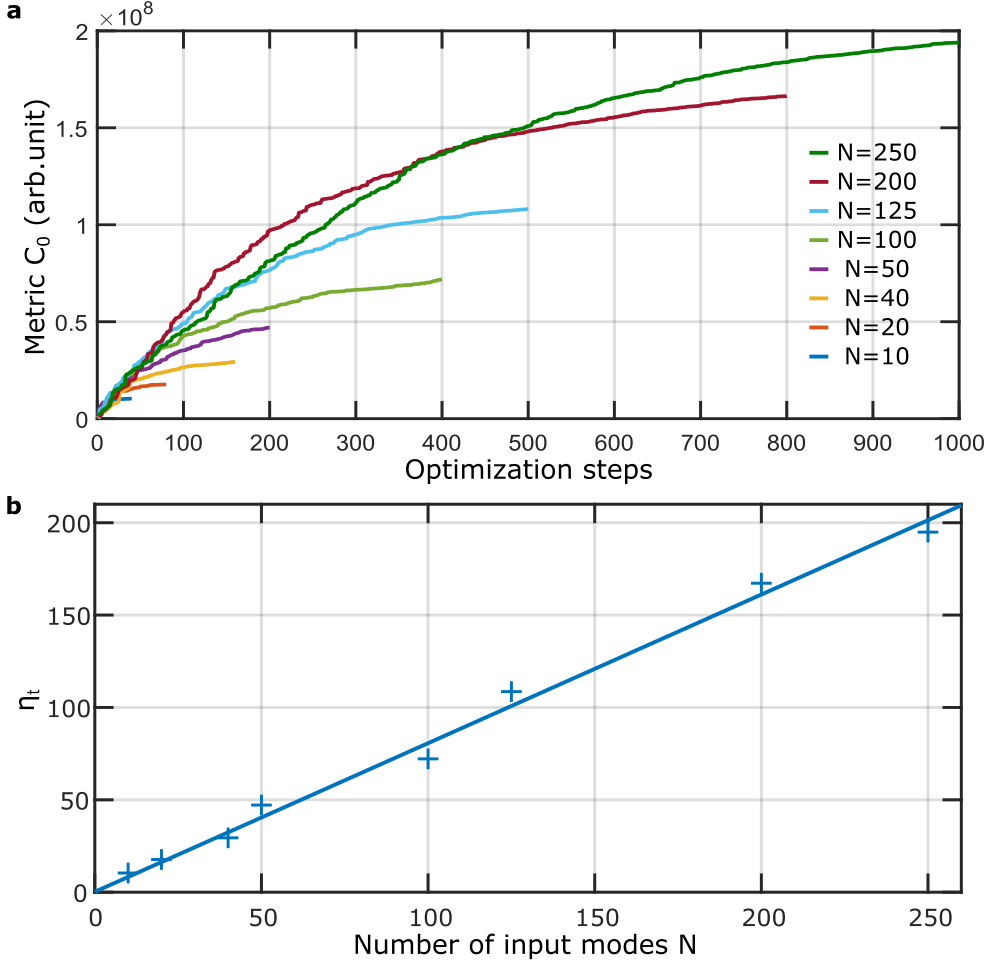}
    \caption{\textbf{Numerical simulations of the optimization process in the single-scattering regime.} \textbf{(a)} Optimization curves of the feedback metric $C_0$ as a function of iteration step for $8$ different numbers of controlled input modes (SLM macropixels): $N = 10$ (dark blue), $N=20$ (orange), $N=40$ (yellow), $N=50$ (purple), $N=100$ (light green), $N=125$ (light blue), $N=200$ (red), and $N=250$ (dark green). Each curve spans $4N$ iterations and is averaged over $10$ independent disorder realizations. \textbf{(b)} Enhancement factor $\eta_t$ extracted from the data in \textbf{(a)} as a function of the number of controlled modes $N$. The solid line represents a linear fit of the form $a(N-1)+1$, demonstrating a proportional dependence on $N-1$ with a slope of $a=0.80$ and $R^2=0.9920$, closely matching the theoretical prediction of $\pi/4$.}
    \label{FigureSM1}
\end{figure}

Figure~\ref{FigureSM1}a shows the simulated optimization curves of the feedback metric $C_0$ for $8$ different numbers of controlled input modes. 
In this context, the number of controlled input modes corresponds to the number of macropixels $N$ utilized on the SLM to perform the random partitioning algorithm. Each optimization curve spans $4N$ iteration steps and represents an average over $10$ independent realizations of the disorder.

Figure~\ref{FigureSM1}b shows the enhancement factor $\eta_t$ obtained as a function of the number of controlled input modes. In excellent agreement with Eq.~(2) of the manuscript, we observe that $\eta_t$ scales linearly with $N-1$. A linear fit to the data yields a slope of $0.80$, which is in close agreement with the theoretical prediction of $\pi/4 \approx 0.78$.

\subsection{Simulations in the multiple-scattering regime}
\label{simulationsthick}

To simplify the computations in the thick scattering medium simulations, we use the intermediate result from Eq.~\eqref{fullqqq}. 
In particular, it can be seen from Eq.~\eqref{fullqqq} that $C_0$ can be calculated by computing the matrix product $M_\alpha V_{\text{SLM}}$, where $M_\alpha$ is an $M \times N$ matrix composed of the coefficients $\{ \alpha_{k0l} \}_{k \in [1,N], l \in [1,M]}$ and $V_{\text{SLM}}$ is a vector containing the SLM phase shifts $\{ e^{i \theta_k} \}_{k \in [1,N]}$, and then summing the squared absolute values of all elements in the resulting vector. 
Furthermore, the coefficients $\alpha_{k0l}$ are generated in the same manner as the scattering matrix coefficients (since they share the same statistical properties) using the MATLAB command \texttt{(randn(M, N) + 1i * randn(M, N)) / sqrt(2)}, where $N$ directly denotes the number of controlled modes.

\begin{figure}[h!]
    \centering
    \includegraphics[width=0.7\linewidth]{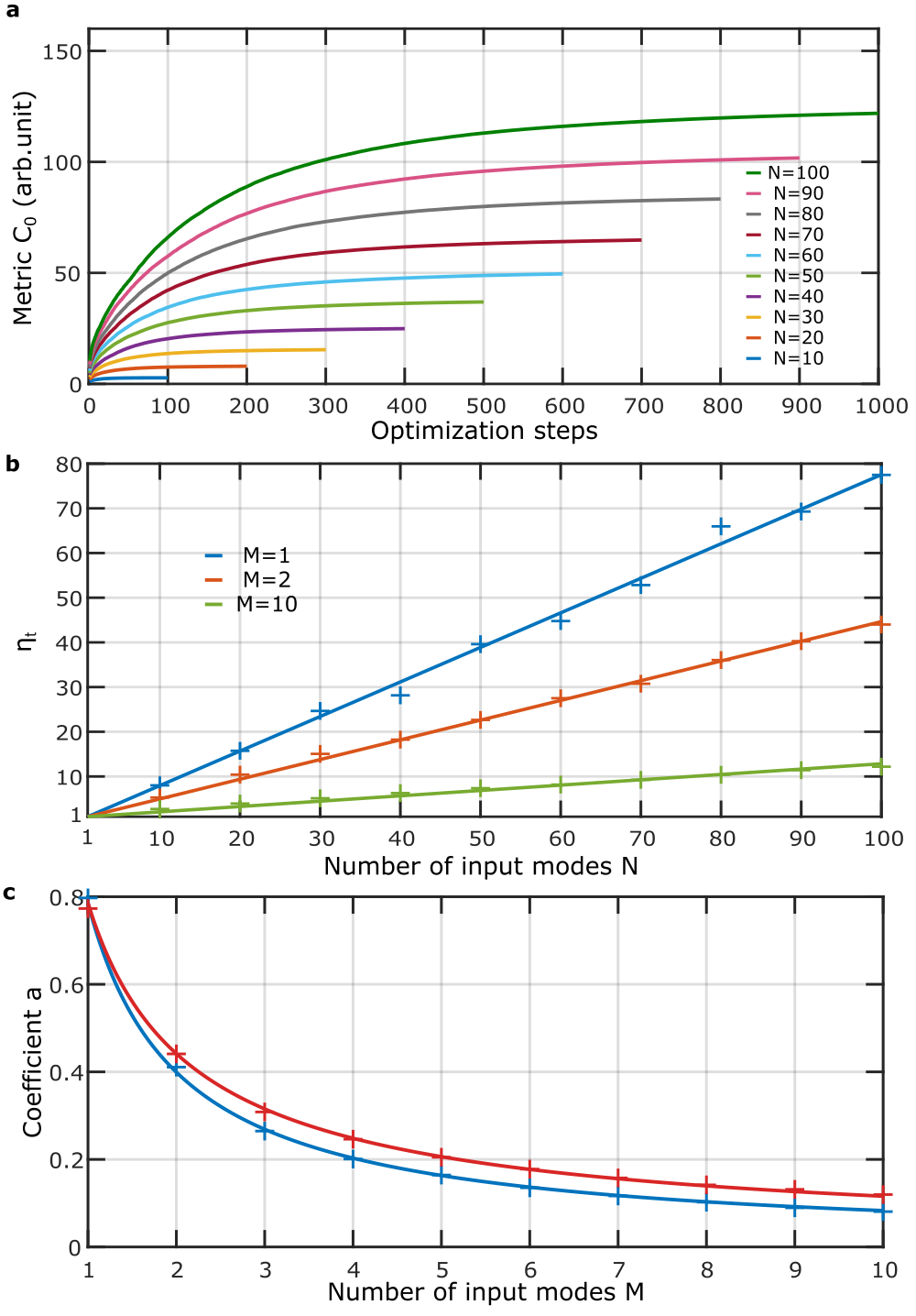}
    \caption{\textbf{Numerical simulations of the optimization process in the multiple-scattering regime.} Optimization curves of the feedback metric $C_0$ as a function of iteration step for $10$ different numbers of controlled input modes (SLM macropixels) $N$, with $M=10$. Each curve spans $10N$ iterations and is averaged over $100$ independent disorder realizations. \textbf{(b)} Enhancement factor $\eta_t$ extracted from the data in \textbf{(a)} as a function of the number of controlled modes $N$ for three different values of $M$: $M=1$ (blue), $M=2$ (orange), and $M=10$ (green). The solid lines represent linear fits of the form $a(N-1)+1$, demonstrating a proportional dependence on $N-1$ with slopes of $a=0.77$, $a=0.44$, and $a=0.12$, respectively ($R^2>0.97$ in all cases). \textbf{(c)} The red markers show the dependence of the slope coefficient $a$ on the number of measured output modes $M$. The solid red line corresponds to a fit of the form $\frac{\pi}{4} M^{-b}$, yielding $b \approx 0.83$ with $R^2=0.999$. The blue markers and corresponding fit represent the same scenario, but where the SLM phase pattern is explicitly constrained to maximize only the single maximum-amplitude term of the sum, corresponding to the initial assumption used in the analytical derivation of Eq.~\eqref{formulahp}. In this constrained case, the fit yields $b \approx 0.98 \approx 1$ with $R^2=0.999$, in excellent agreement with the analytical model. }
    \label{FigureSM2}
\end{figure}

Figure~\ref{FigureSM2}a shows the simulated optimization curves of the feedback metric $C_0$ for $10$ different numbers of controlled input modes $N$, with a fixed number of output modes $M=10$. Each optimization curve spans $10N$ iteration steps and represents an average over $100$ independent realizations of the disorder.

Figure~\ref{FigureSM2}b presents the enhancement factor $\eta_t$ obtained as a function of the number of controlled input modes $N$ for three different values of $M$ ($1$, $2$, and $10$). In excellent agreement with Eq.~(3) of the manuscript, we observe that $\eta_t$ scales linearly with $N-1$ in all cases. A linear fit to the data yields a slope of $0.77$ for $M=1$, $0.44$ for $M=2$, and $0.12$ for $M=10$.

The red markers in Figure~\ref{FigureSM2}c show the dependence of the slope coefficients from Figure~\ref{FigureSM2}b as a function of the number of spatial modes $M$. 
A fit of the form $\frac{\pi}{4} M^{-b}$ yields a value of $b \approx 0.83$ ($R^2=0.999$), which differs from the strict $1/M$ dependence predicted by Eq.~\eqref{formulahp}. 
As mentioned in Section~\ref{equation3section}, this discrepancy arises because the optimal phase pattern found by the algorithm is not the one that maximizes the single term of largest amplitude in the sum - an assumption that was initially used to derive Eq.~\eqref{formulahp}. 

To verify that this assumption is really the source of the deviation, we performed complementary simulations in which the SLM was explicitly programmed to maximize only the contribution of the single brightest term. 
In this constrained scenario, we recover the theoretical $1/M$ dependence, as shown by the blue curve in Figure~\ref{FigureSM2}c. 
We can thus conclude that the phase pattern optimizing $C_0$ is not the one that maximizes the largest term of the sum. 
Instead, the solution found by the random partitioning optimization achieves a more complex balance across multiple terms, leading to the $1/M^b$ scaling with $b \approx 0.83$. 
Notably, because $b < 1$, the optimal phase pattern found by the algorithm actually performs better, yielding a higher net enhancement than what would be expected under the single-term maximization hypothesis.

\section{Additional experimental details}

\subsection{Details on the experimental setup used to obtain the results shown in Figures 1 and 2 of the manuscript}

To obtain the results presented in Figs.~1 and 2 of the main text, we used the experimental setup depicted in Fig.~1(a) of the manuscript with the following specific components:

\begin{itemize}
    \item \textbf{SPAD camera:} The data were acquired using a time-gated SPAD array (Hermes, Micro Photon Devices). It has a resolution of $64 \times 32$ pixels with a $150~\mu$m pixel pitch and operates at a maximum frame rate of 96~kfps. Its photon detection efficiency at 810~nm is approximately $5\%$. For our measurements, the exposure time was fixed at $1~\mu$s.
    \item \textbf{SLM:} We used a phase-only SLM (Pluto, Holoeye).
    \item \textbf{Nonlinear crystal:} Photon pairs were generated using a $1 \times 5 \times 5$~mm$^3$ type-II $\beta$-barium borate (BBO) crystal (Newlight Photonics).
    \item \textbf{Pump source:} The crystal was pumped by a continuous-wave blue laser diode (LBX-405, Oxxius) delivering 100~mW of output power.
\end{itemize}

\subsection{Details on the experimental setup used to obtain the results shown in Figure 3 of the manuscript}

\begin{figure}[h!]
    \centering
    \includegraphics[width=0.8\linewidth]{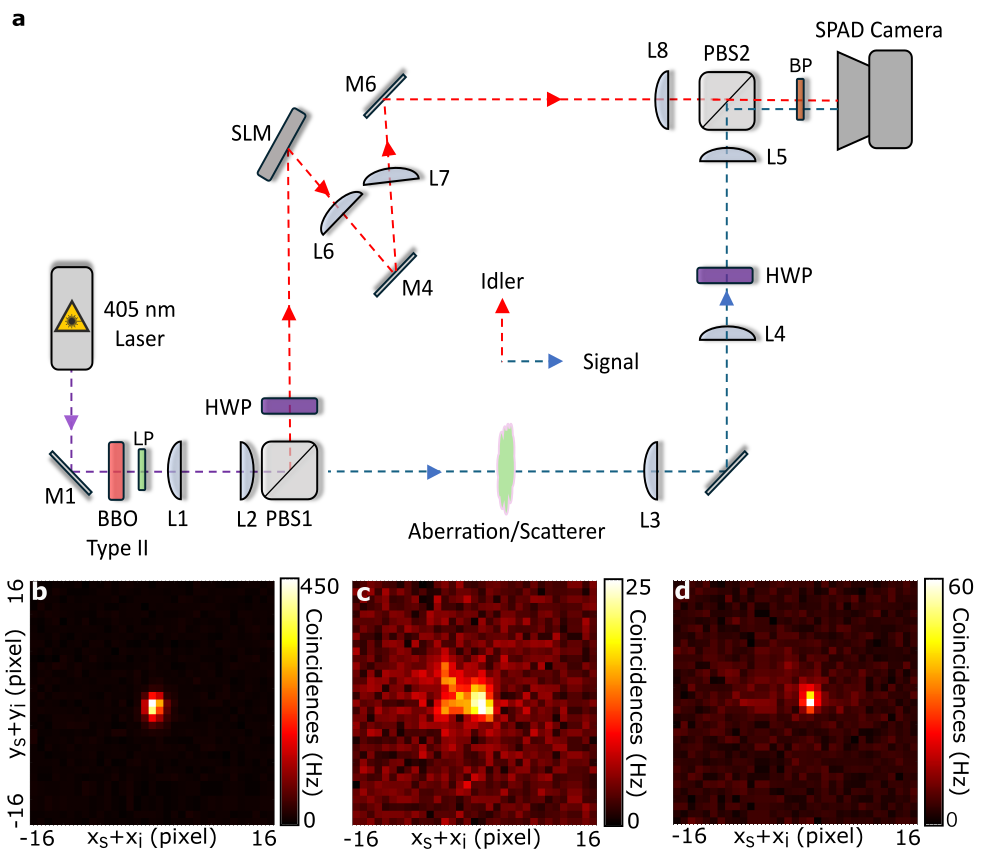}
    \caption{\textbf{Experimental setup and results for the single-scattering regime.} \textbf{(a)} Experimental setup used to obtain the results shown in Fig.~3 of the main text. This configuration is a slightly modified version of the setup depicted in Fig.~1a of the manuscript. In the idler arm, an arrangement of five lenses ({$L_1=40$~mm, $L_2=200$~mm, $L_6=100$~mm, $L_7=100$~mm, and $L_8=100$~mm}) projects the Fourier transform of the crystal surface onto one half of the SPAD camera sensor. In the signal arm, a similar five-lens arrangement ({$L_1$, $L_2$, $L_3=100$~mm, $L_4=100$~mm, and $L_5=100$~mm}) projects the Fourier transform of the crystal surface onto the other half of the sensor. The camera used in this specific configuration is a SPAD512 model (PiImaging). \textbf{(b)} Sum-coordinate projection of $G^{(2)}$ acquired without scattering medium.\textbf{(c)} Sum-coordinate projection of $G^{(2)}$ acquired in the presence of the Parafilm layer with a flat phase applied to the SLM. \textbf{(d)} Sum-coordinate projection of $G^{(2)}$ acquired with the optimized phase mask displayed on the SLM.}
    \label{FigureSM3}
\end{figure}

To obtain the results presented in Fig.~3 of the manuscript, several modifications were made to the experimental setup compared to the configuration used for Figs.~1 and 2:

\begin{itemize}
    \item \textbf{SPAD camera:} We replaced the original SPAD array (Hermes, Micro Photon Devices) with an alternative time-gated SPAD camera (SPAD512, Pi Imaging) due to the temporary unavailability of the Hermes sensor during this measurement campaign. The SPAD512 features a spatial resolution of $512 \times 512$ pixels with a $16.38~\mu$m pixel pitch, a maximum frame rate of 100~kfps, and a photon detection efficiency of approximately $13\%$ at 810~nm. For these experiments, the exposure time was set to $7~\mu$s.
    \item \textbf{Pump source:} Due to a technical issue with the original diode, the pump source was substituted with an identical model (LBX-405, Oxxius) operating at a reduced output power of 50~mW.
    \item \textbf{Optical adjustments:} As detailed in the schematic in Fig.~\ref{FigureSM3}a, minor adjustments to the optical setup (e.g. lens substitutions and mirror repositioning) were made to adapt the system's magnification to the new camera's specifications. As a result, the $G^{(2)}$ sum-coordinate projection measured without the scattering medium (Fig.~\ref{FigureSM3}b) also exhibits a strong and sharp correlation peak. 
\end{itemize}

To provide a complete view of the measurement sequence (without the medium, with the medium, and after correction), we also include Figs.~\ref{FigureSM3}c and \ref{FigureSM3}d, which are identical to Figs.~3a and 3c of the main manuscript.

\clearpage

\bibliography{Biblio2}
\end{document}